\documentclass[a4paper,11pt]{article}
\usepackage{pos}
\usepackage{orcidlink}

\title{Symmetric Mass Generation via Multicriticality in a 3D Lattice Gross–Neveu Model}

\author[a]{Sandip Maiti~\orcidlink{0000-0002-5248-5316}}
\author*[a,b]{Debasish Banerjee~\orcidlink{0000-0003-0244-4337}}
\author[c]{Shailesh Chandrasekharan~\orcidlink{0000-0002-3711-4998}}
\author[d]{Marina K. Marinkovic~\orcidlink{0000-0002-9883-7866}}

\affiliation[a]{Saha Institute of Nuclear Physics, HBNI, 1/AF Bidhannagar, Kolkata 700064, India}

\affiliation[b]{School of Physics and Astronomy, University of Southampton, University Road, SO17 1BJ, UK}

\affiliation[c]{Department of Physics, Box 90305, Duke University, Durham, North Carolina 27708, USA}

\affiliation[d]{Institut für Theoretische Physik, Wolfgang-Pauli-Straße 27, ETH Zürich, 8093 Zürich, Switzerland}

\emailAdd{D.Banerjee@soton.ac.uk}

\abstract{
We investigate a three-dimensional lattice model of two flavors of massless staggered fermions
coupled through two independent four-fermion interactions, $U_I$ and $U_B$. Using large-scale
fermion-bag Monte Carlo simulations, we map out the phase diagram in the $(U_I, U_B)$ parameter
space and identify three distinct phases: a massless fermion phase, a symmetry-broken massive
phase, and a symmetric massive phase. When one of the interactions is absent ($U_B=0$), the
system undergoes a single continuous transition directly connecting the massless and symmetric
massive phases, a feature previously associated with unconventional fermion mass generation.
We find that turning on a nonzero $U_B$ separates this direct transition into two successive transitions with an intermediate symmetry-broken phase. The transition from the massless to the broken phase belongs to the Gross-Neveu universality class, while the transition from the broken to the symmetric massive phase falls into the three-dimensional XY universality class. Our results indicate that the special point at vanishing coupling, where the direct transition occurs, plays the role of a multicritical point organizing the surrounding phase structure. These findings provide a unified lattice perspective on conventional and unconventional mechanisms of fermion mass generation within a single model.}

\FullConference{The 42nd International Symposium on Lattice Field Theory (LATTICE2025)\\
2-8 November 2025\\
Tata Institute of Fundamental Research, Mumbai, India\\}

\begin{document}
\maketitle

\section{Introduction}
Mass generation for fermions is typically understood as a consequence of either explicit 
fermion bilinear mass terms present in the action or dynamical symmetry breaking
where a continuous global symmetry spontaneously breaks (SSB) \cite{Gross:1974jv}. 
This paradigm underlies the Higgs mechanism in the Standard Model and chiral symmetry 
breaking in Quantum Chromodynamics, where strong interactions induce fermion condensates 
and generate mass scales. These examples have shaped the widespread view that symmetry 
breaking is essential for fermion mass generation.

Recent lattice studies have challenged this expectation by demonstrating that fermions can become massive without forming bilinear condensates and without breaking any symmetries of the massless theory \cite{Slagle:2014vma, Tong:2021phe}. In this scenario, known as symmetric mass generation (SMG), masses arise purely from strong interactions \cite{Wang:2022ucy}. Interest in SMG has grown following the observation of continuous lattice transitions directly connecting massless and symmetric massive phases, suggesting the presence of a novel critical point governing the infrared physics. Evidence for such behavior has been found in three-dimensional lattice models \cite{Ayyar:2014eua}, where, suprisingly, the tuning of a single coupling was sufficient to make the system critical, hinting that the symmetry of the lattice model was playing a role.

In this work, we extend the three-dimensional lattice model of massless staggered fermions studied earlier in Ref.~\cite{Ayyar:2014eua}, by introducing an additional four-fermion interaction that breaks part of the lattice symmetry \cite{Ayyar:2014eua}. Using large-scale Monte Carlo simulations based on the fermion-bag formulation, we explore how this interaction  modifies the phase structure. We find that the direct massless-to-SMG transition is replaced by two distinct continuous transitions separated by an intermediate phase with conventional symmetry breaking \cite{Maiti:2025ley}. The transition from the massless phase to the conventional spontaneously symmetry broken (SSB) phase is consistent with Gross–Neveu criticality, while the transition between the SSB and symmetric massive phases is governed by a bosonic universality class consistent with three-dimensional XY behavior, reflecting the decoupling of massive fermions. This suggests that the direct transition observed in \cite{Ayyar:2014eua} is in fact a multicritical point which was accessible easily only due to the enhanced symmetry of the lattice model.

\section{Model and Symmetries}
 Our microscopic model is defined on a three-dimensional Euclidean space--time lattice, describing two flavors of massless staggered fermions, labeled $u$ and $d$. The fermions interact through two independent four-fermion couplings, denoted $U_I$ and $U_B$. At each lattice site $x$, the theory contains four Grassmann-valued fields, $\bar u_x$, $u_x$, $\bar d_x$, and $d_x$. The full lattice action takes the form
\begin{align}
S
= \sum_{x,y} \bigl(\bar u_x M_{xy} u_y + \bar d_x M_{xy} d_y\bigr)
\;-\; U_I \sum_{x} \bigl( \bar u_x u_x \,\bar d_x d_x \bigr)
\;-\; U_B \sum_{\langle x,y\rangle} 
\bigl( \bar u_x u_x\, \bar u_y u_y 
      + \bar d_x d_x\, \bar d_y d_y \bigr).
\label{eq:model}
\end{align}

Here $x$ and $y$ label sites on a three-dimensional cubic lattice. Lattice sites are represented by vectors $x=(x_0,x_1,x_2)$ with integer coordinates $x_\mu=0,1,\ldots,L-1$, so that the total lattice volume is $V = L^3$. Nearest-neighbor sites are related by shifts along the coordinate directions $x\to x\pm\hat\mu$, where $\hat\mu$ ($\mu=0,1,2$) denote unit vectors. 
The matrix $M$ appearing in Eq.~\eqref{eq:model} is the massless staggered fermion operator, defined by
\begin{align}
M_{xy}
\;=\;
\frac{1}{2}\sum_{\hat\mu}
\eta_{x,\hat\mu}
\left(
\delta_{x+\hat\mu,y}
-
\delta_{x-\hat\mu,y}
\right),
\end{align}
which is antisymmetric and connects only nearest-neighbor lattice sites.
The staggered phase factors $\eta_{x,\hat\mu}$ are chosen to implement a $\pi$ flux through each elementary plaquette of the lattice. In the convention used here, they are given by
\begin{equation}
\eta_{x,\hat0}=1, \qquad
\eta_{x,\hat1}=(-1)^{x_0}, \qquad
\eta_{x,\hat2}=(-1)^{x_0+x_1}.
\end{equation}
Finally, lattice sites are classified as even or odd according to the parity of
$x_0+x_1+x_2$.

The action in Eq.~\eqref{eq:model} contains three terms with distinct symmetry properties. The nearest-neighbor hopping term describes free staggered fermions and is invariant under a global $SU(4)\times U(1)$ symmetry. On even sites $x$ and odd sites $y$, the fermion fields transform as
\begin{align}
\begin{pmatrix}
u_x \\
\bar u_x \\
d_x \\
\bar d_x
\end{pmatrix}
&\;\to\;
V\, e^{i\theta}
\begin{pmatrix}
u_x \\
\bar u_x \\
d_x \\
\bar d_x
\end{pmatrix},
\qquad
\begin{pmatrix}
\bar u_y \\
u_y \\
\bar d_y \\
d_y
\end{pmatrix}
\;\to\;
V^* e^{-i\theta}
\begin{pmatrix}
\bar u_y \\
u_y \\
\bar d_y \\
d_y
\end{pmatrix},
\label{eq:symm-1}
\end{align}
with $V\in SU(4)$.

The on-site interaction proportional to $U_I$ preserves the $SU(4)$ symmetry but breaks the axial $U(1)$ symmetry, while the bond interaction proportional to $U_B$ is invariant under independent $SU(2)\times U(1)$ transformations acting on each flavor. For the $u$ fermions, this symmetry is given by
\begin{align}
\begin{pmatrix}
u_x \\
\bar u_x
\end{pmatrix}
&\;\to\;
V_u\, e^{i\theta_u}
\begin{pmatrix}
u_x \\
\bar u_x
\end{pmatrix},
\qquad
\begin{pmatrix}
\bar u_y \\
u_y
\end{pmatrix}
\;\to\;
V_u^* e^{-i\theta_u}
\begin{pmatrix}
\bar u_y \\
u_y
\end{pmatrix},
\label{eq:symm-2}
\end{align}
with $V_u\in SU(2)$, and an analogous transformation for the $d$ fermions.

When $U_I\neq0$ and $U_B=0$, the model retains the full $SU(4)$ symmetry. For nonzero $U_I$ and $U_B$, the symmetry is reduced to $SU(2)\times SU(2)\times U_\chi(1)$, where $U_\chi(1)$ corresponds to the axial combination $\theta_u=-\theta_d$.

The spontaneous breaking of the $U_\chi(1)$ symmetry is associated with the development of fermion bilinear order parameters, $\langle \bar u u\rangle$ and $\langle \bar d d\rangle$. Instead of measuring these quantities directly, we characterize the symmetry-breaking behavior through correlation functions of local fermion bilinears. To this end, we define the following susceptibilities:
\begin{equation}
\chi_{ud} = \frac{1}{2V}\sum_{x,y}
\big\langle \bar u_x u_x \,\bar d_y d_y \big\rangle,\qquad
\chi_{uu} = \frac{1}{2V}\sum_{x,y}
\big\langle \bar u_x u_x \,\bar u_y u_y \big\rangle,\qquad
\chi_{dd} = \frac{1}{2V}\sum_{x,y}
\big\langle \bar d_x d_x \,\bar d_y d_y \big\rangle.
\label{eq:obs-chi}
\end{equation}
Expectation values are taken with respect to the path integral defined by the action in Eq.~\eqref{eq:model}. Flavor symmetry implies $\chi_{uu}=\chi_{dd}$, and we find numerically that $\chi_{ud}$ is approximately equal to $\chi_{uu}$ on sufficiently large system sizes and large values of $U_I$.

In phases with fermion bilinear order, these susceptibilities grow extensively with system size, scaling as $\chi \sim V$. By contrast, in both the massless fermion phase and the symmetric massive phase, they remain finite in the thermodynamic limit. At a continuous phase transition, finite-size scaling predicts the behavior $\chi \sim L^{2-\eta}$, where $\eta$ denotes the anomalous dimension of the corresponding bilinear operator.

\section{Fermion Bag Method}
\label{sec-3}

To compute the susceptibilities defined in Eqs.~\eqref{eq:obs-chi}, we employ the fermion bag formulation \cite{Chandrasekharan:2009wc}, which provides a sign-problem-free Monte Carlo representation of the path integral. Expectation values of Grassmann-valued observables $\hat O$ are defined as
\begin{align}
\langle \hat O \rangle
=
\frac{1}{Z}
\int [\mathcal{D}\bar d\, \mathcal{D} d\, \mathcal{D}\bar u\, \mathcal{D} u]\;
O\, e^{-S},
\end{align}
with the partition function $Z$ defined analogously and the action $S$ given in Eq.~\eqref{eq:model}.

\begin{figure}
    \centering
    \includegraphics[width=0.6\linewidth, trim=100 560 250 100, clip]{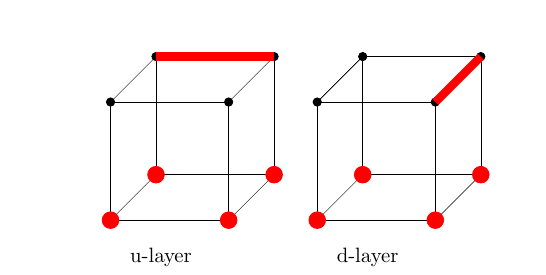}
    \caption{Example of a $[b,i]$ configuration on a $2^3$ lattice. Red circles show instantons, red links show bonds or dimers, and black circles indicate free sites.}
    \label{fig:fbag}
\end{figure}

The four-fermion interactions are encoded using discrete auxiliary variables:
nearest-neighbor \emph{bond variables} $b_{xy}$ and on-site \emph{instanton variables} $i_x$.
A given configuration $[b,i]$ partitions the lattice into disconnected regions,
referred to as \emph{fermion bags}, within which the Grassmann integrals can be
evaluated exactly. Figure~\ref{fig:fbag} shows an example configuration on an $L=2$ lattice. The essential idea of the fermion bag formulation is to reorganize the Grassmann path integral as a sum over such bag configurations. Depending on how the interaction terms are expanded, this reorganization admits two complementary descriptions, commonly referred to as the strong-coupling and weak-coupling viewpoints.

\vspace{0.6em}
\textbf{(1) Strong-coupling viewpoint}

In the strong-coupling formulation, the interaction terms are expanded locally as
\begin{align}
e^{U_B\,\bar u_x u_x\,\bar u_y u_y}
&= \sum_{b_{xy}=0,1}
\left(U_B\,\bar u_x u_x\,\bar u_y u_y\right)^{b_{xy}}, \\
e^{U_I\,\bar u_x u_x\,\bar d_x d_x}
&= \sum_{i_x=0,1}
\left(U_I\,\bar u_x u_x\,\bar d_x d_x\right)^{i_x}.
\end{align}
After integrating over Grassmann variables on sites occupied by bonds or instantons, the partition function becomes
\begin{equation}
Z = \sum_{[b,i]} (U_I)^{N_i} (U_B)^{N_u+N_d}
\, \det(W_u)\, \det(W_d),
\end{equation}
where $W_u$ and $W_d$ are free staggered fermion matrices restricted to the
unoccupied lattice sites.
Connected clusters of these sites define \emph{strong-coupling fermion bags}.

\vspace{0.8em}
\textbf{(2) Weak-coupling viewpoint}

In the weak-coupling formulation, the interactions are treated as operator insertions in an expansion about the free staggered fermion theory. The partition function can then be expressed schematically as
\begin{equation}
Z = \sum_{[b,i]} (U_I)^{N_i} (U_B)^{N_u+N_d}
\langle \bar u_{a_1} u_{a_1}\cdots \bar u_{a_l} u_{a_l} \rangle
\langle \bar d_{b_1} d_{b_1}\cdots \bar d_{b_k} d_{b_k} \rangle ,
\end{equation}
where $\{a_1,\ldots,a_l\}$ and $\{b_1,\ldots,b_k\}$ denote the lattice sites at
which interaction insertions occur in the $u$ and $d$ fermion sectors,
respectively.

Applying Wick’s theorem, these correlation functions reduce to determinants of
matrices built from free staggered fermion propagators between the insertion
sites. The partition function then takes the form
\begin{equation}
Z = (\det M)^2
\sum_{[b,i]} (U_I)^{N_i} (U_B)^{N_u+N_d}
\, \det(G_u)\, \det(G_d),
\end{equation}
where $G_u$ and $G_d$ are propagator matrices constructed from $M^{-1}$ between
the corresponding interaction sites.
In this representation, fermion bags are identified with connected clusters of
interaction sites and are referred to as \emph{weak-coupling fermion bags}.

\section{Monte Carlo Implementation and Thermalization}
\label{sec-MC}

We simulate the fermion bag representation using a Markov-chain Monte Carlo algorithm that samples bond and instanton configurations $[b,i]$. The detailed update procedure and determinant evaluation strategy are described in Ref.~\cite{Maiti:2026log}.

A computationally demanding step in the algorithm is the evaluation of ratios of fermion determinants that appear in the Metropolis accept/reject probabilities. As discussed in Ref.~\cite{Maiti:2026log}, local fluctuations of fermion bags can be expressed as the determinant of a small matrix, which we call the fluctuation matrix. The dimension of this matrix is determined by the number of local insertions added to or removed from a chosen reference (base) fermion bag configuration. Its elements are constructed using the inverse of a background matrix associated with that base configuration. Although this approach eliminates the need to repeatedly compute large determinants, the construction of the fluctuation matrix and the inversion of the background matrix remain the dominant computational costs.

\begin{figure}
    \centering
    \includegraphics[width=0.32\linewidth, trim=0 0 0 0, clip]{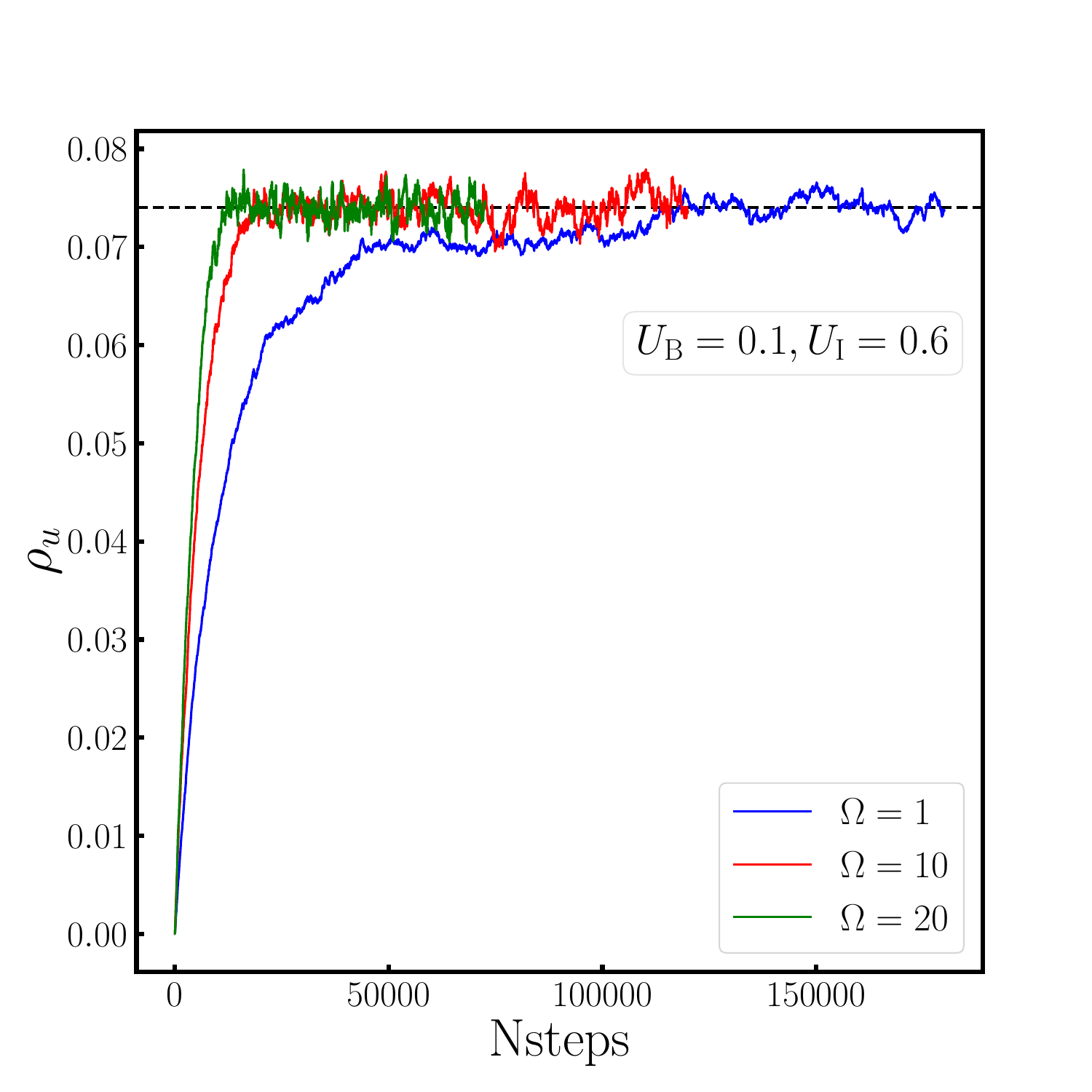}
    \includegraphics[width=0.32\linewidth, trim=0 0 0 0, clip]{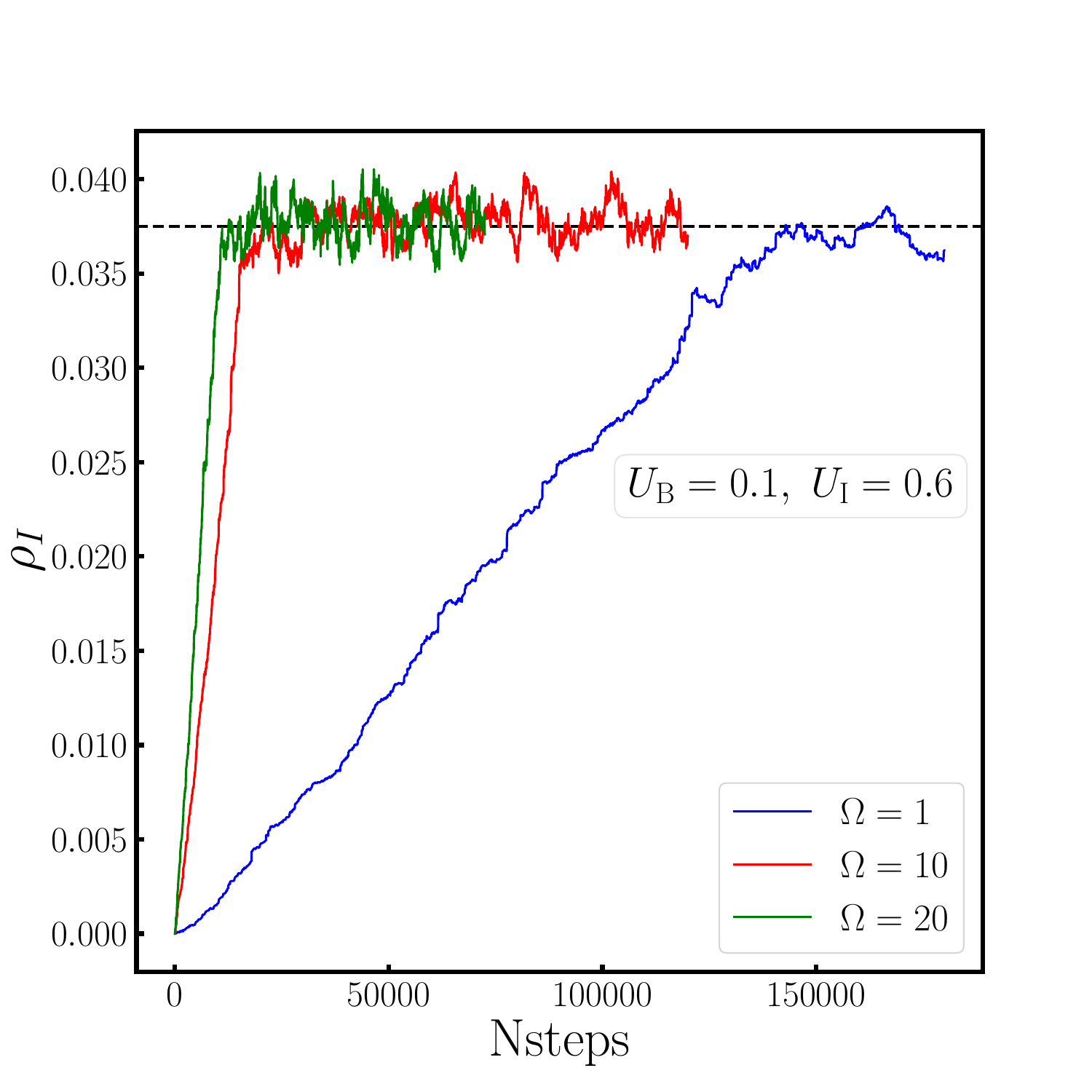}
    \includegraphics[width=0.32\linewidth, trim=0 0 0 0, clip]{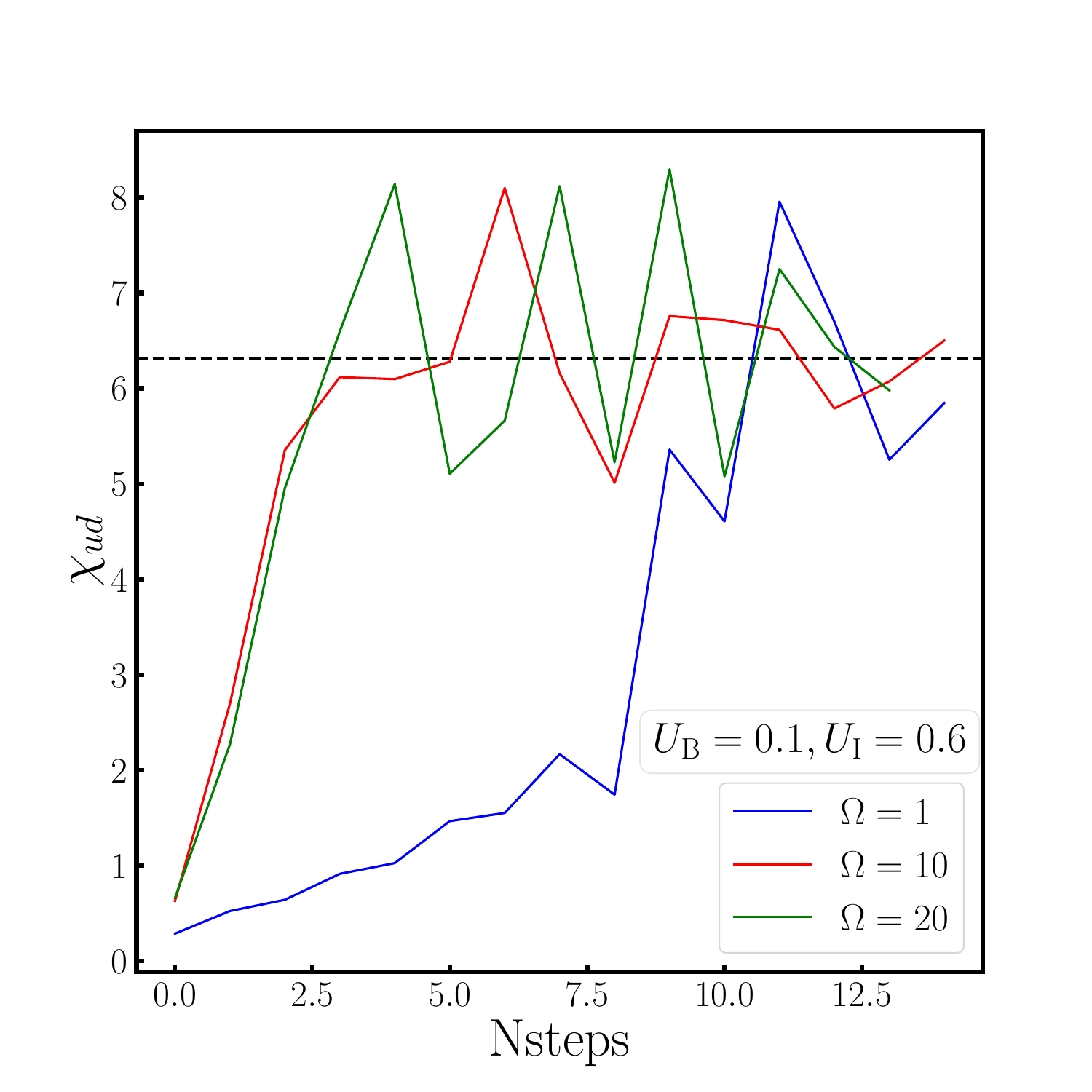}
    \caption{Thermalization histories of the bond density ($\rho_u$) (left), the instanton density ($\rho_I$) (middle), and the susceptibility ($\chi_{ud}$) (right) as functions of Monte Carlo steps for three different values of the reweighting parameter $\Omega = 1, 10,$ and $20$, on a $44^3$ lattice at $U_B=0.1$ and $U_I=0.6$.}
    \label{fig:L44therm}
\end{figure}

Thermalization becomes particularly demanding near criticality, where large fermion bags form and autocorrelation times increase. To improve sampling in regions with broad bag-size distributions, we introduce a reweighting parameter $\Omega$ that enhances the worm sector that contributes to the susceptibility. Figure~\ref{fig:L44therm} shows the evolution of instanton density, bond density, and susceptibilities on an $L=44$ lattice at $U_B=0.1$ and $U_I=0.6$ for $\Omega=1,10,$ and $20$, starting from identical initial configurations and random seeds.

Increasing $\Omega$ accelerates equilibration in terms of Monte Carlo steps—the $\Omega=20$ run reaches equilibrium in slightly fewer steps than $\Omega=10$ but substantially increases the cost per step. While $\Omega=1$ requires $\sim 8$ hours for $1.8\times10^5$ steps, about $7\times10^4$ steps take $\sim 43$ and $\sim 88$ hours for $\Omega=10$ and $20$, respectively. In all cases, the dominant cost arises from constructing the fluctuation matrix. For larger lattices ($L\geq 60$), the rapid growth in the size of the fermion bag and memory requirement makes simulations near criticality prohibitively slow.

\section{The Phase Diagram}
\label{sec-3}
Having established the numerical framework, we now turn to the phase structure of the model in the ($U_I, U_B$) plane. Earlier studies have shown that with only the $U_B$ coupling present, the system exhibits a continuous transition from a massless fermion phase to a phase with spontaneous breaking of the $U_\chi(1)$ symmetry \cite{Chandrasekharan:2011mn}. In contrast, with only the $U_I$ coupling, one finds a direct second-order transition between the massless phase and the SMG phase \cite{Ayyar:2014eua}.

\begin{figure}[t]
    \centering
    \includegraphics[width=0.41\linewidth, trim=10 0 10 0, clip]{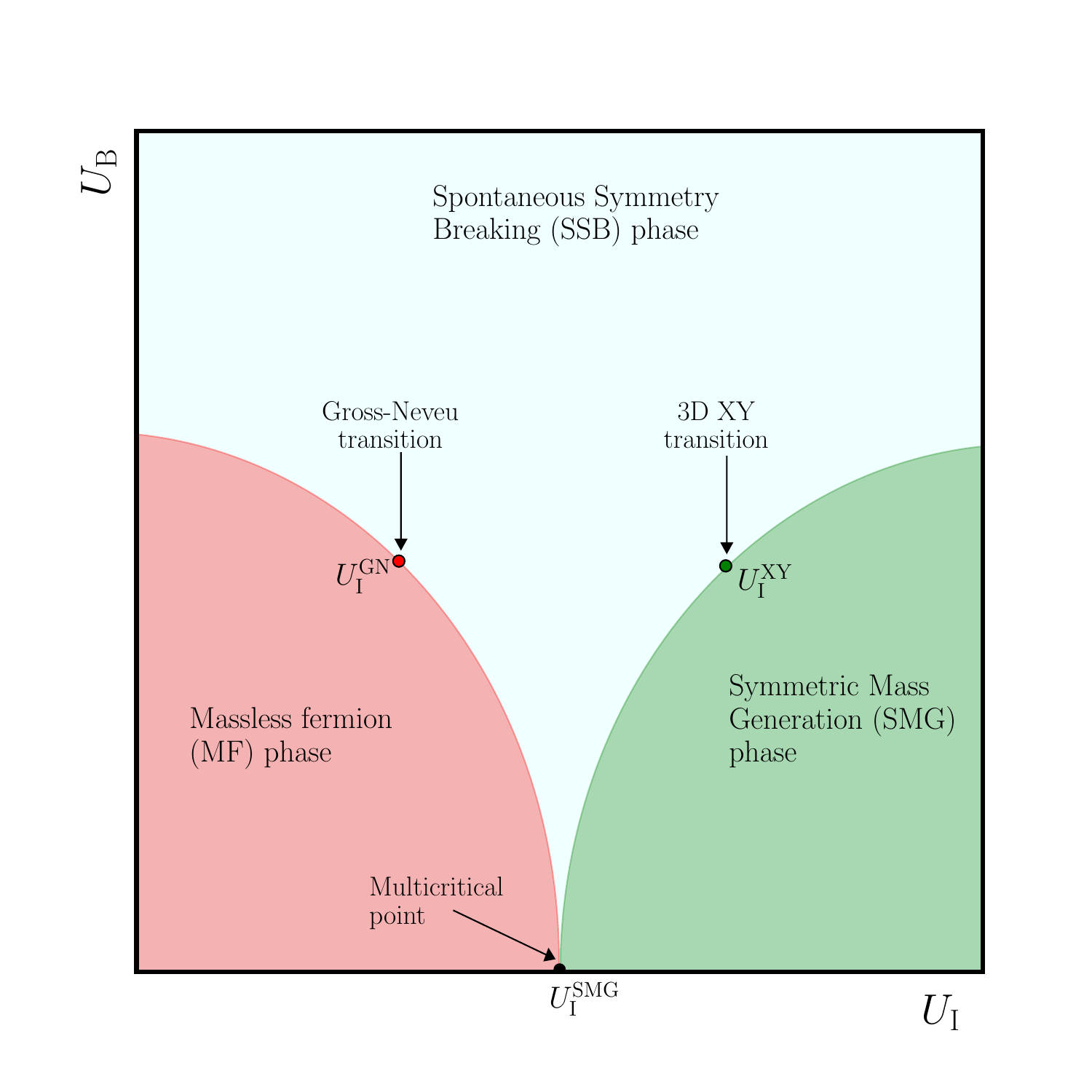}
    \caption{The phase diagram of our model in the $U_I$ and $U_B$ plane.}
    \label{fig:phasedg}
\end{figure}

Our goal is to determine how this picture is modified when a small but nonzero $U_B$ is introduced in the presence of $U_I$. In particular, we ask whether the direct massless–SMG transition splits into two separate transitions separated by an intermediate SSB phase. A schematic phase diagram summarizing this expectation is shown in
Fig.~\ref{fig:phasedg}.
\begin{figure}
    \centering
    \includegraphics[width=0.9\linewidth, trim=200 0 150 0, clip]{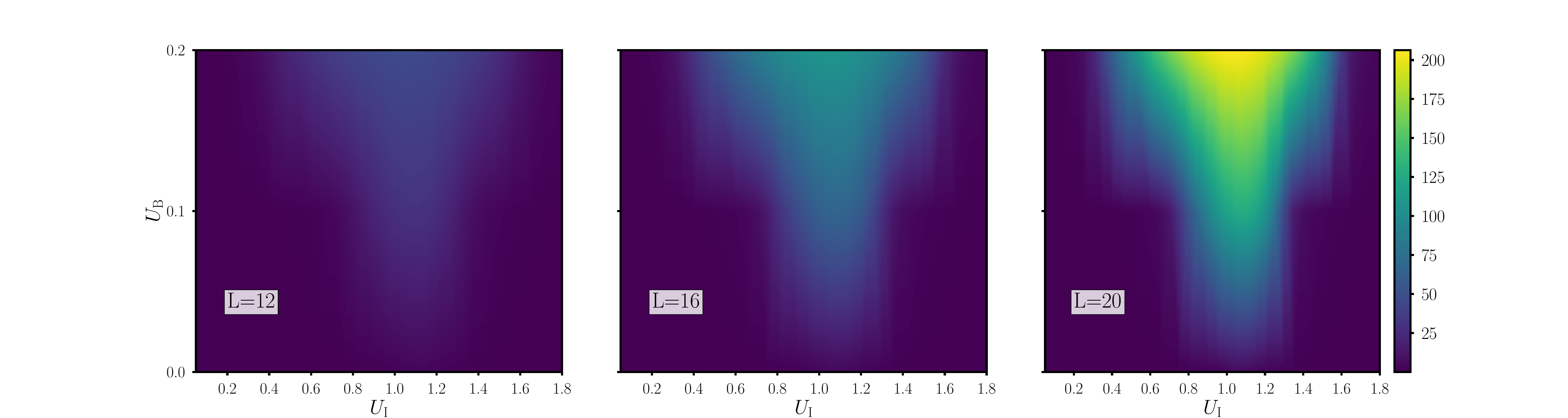}
    \includegraphics[width=0.3\linewidth, trim=0 0 0 0, clip]{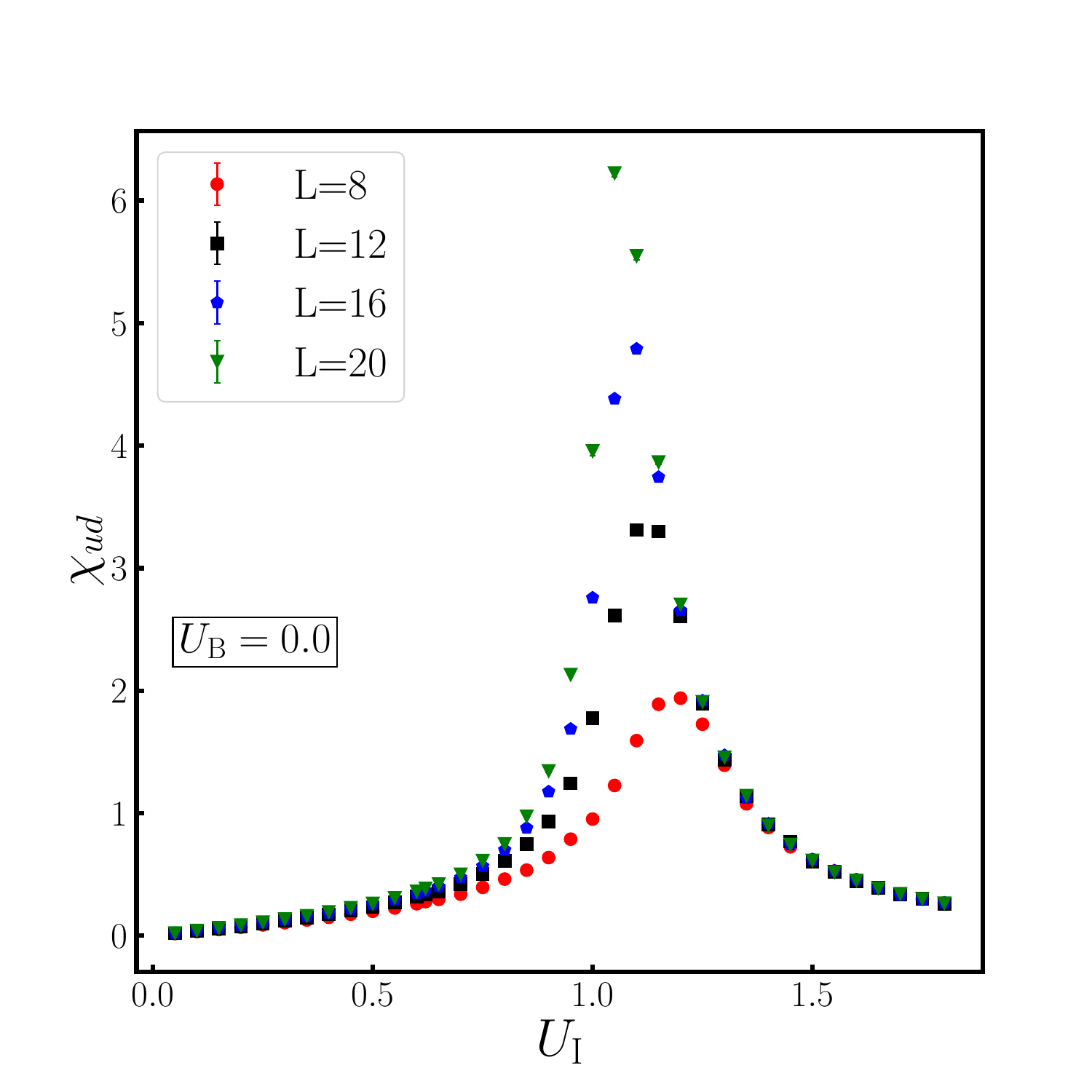}
    \includegraphics[width=0.3\linewidth, trim=0 0 0 0, clip]{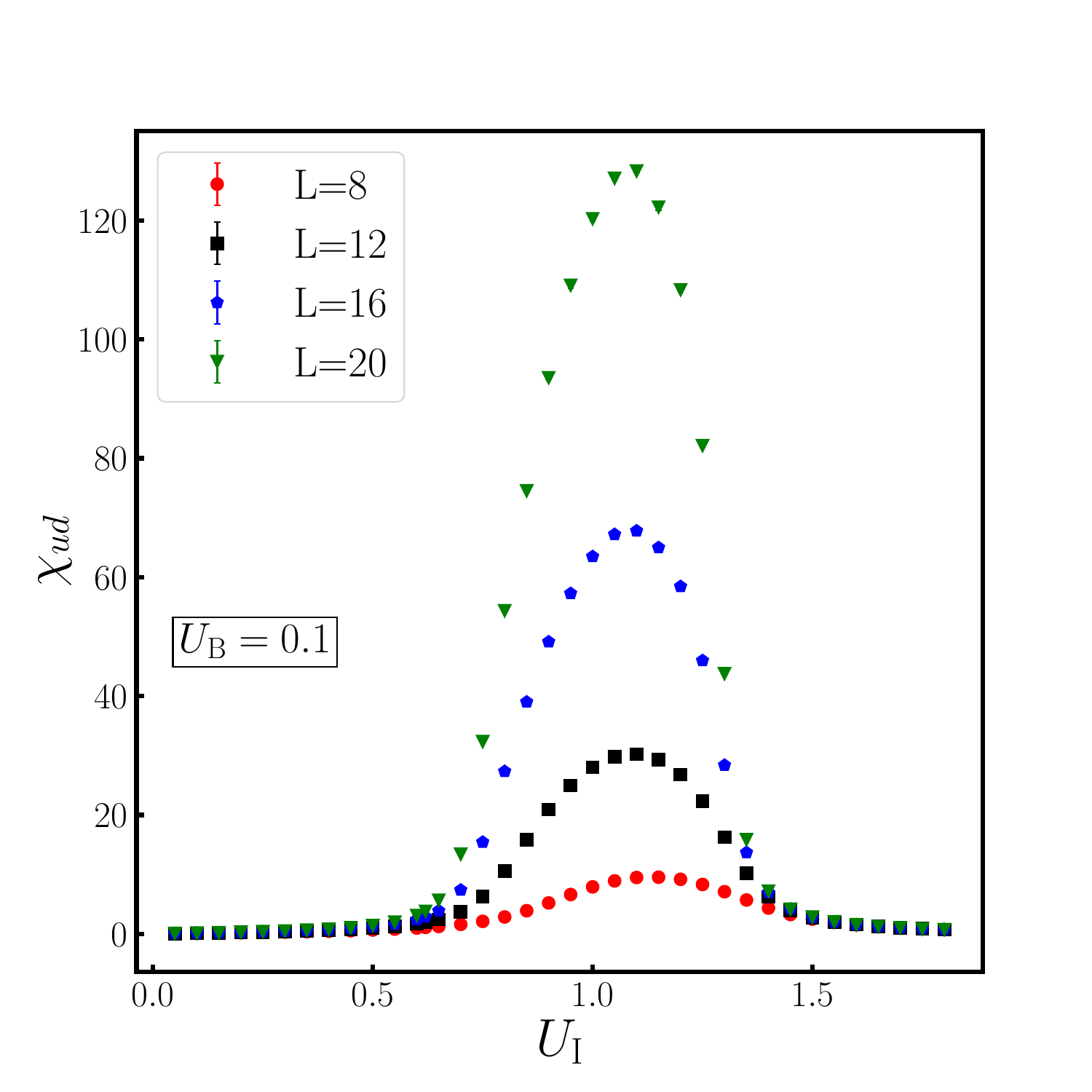}
    \includegraphics[width=0.3\linewidth, trim=0 0 0 0, clip]{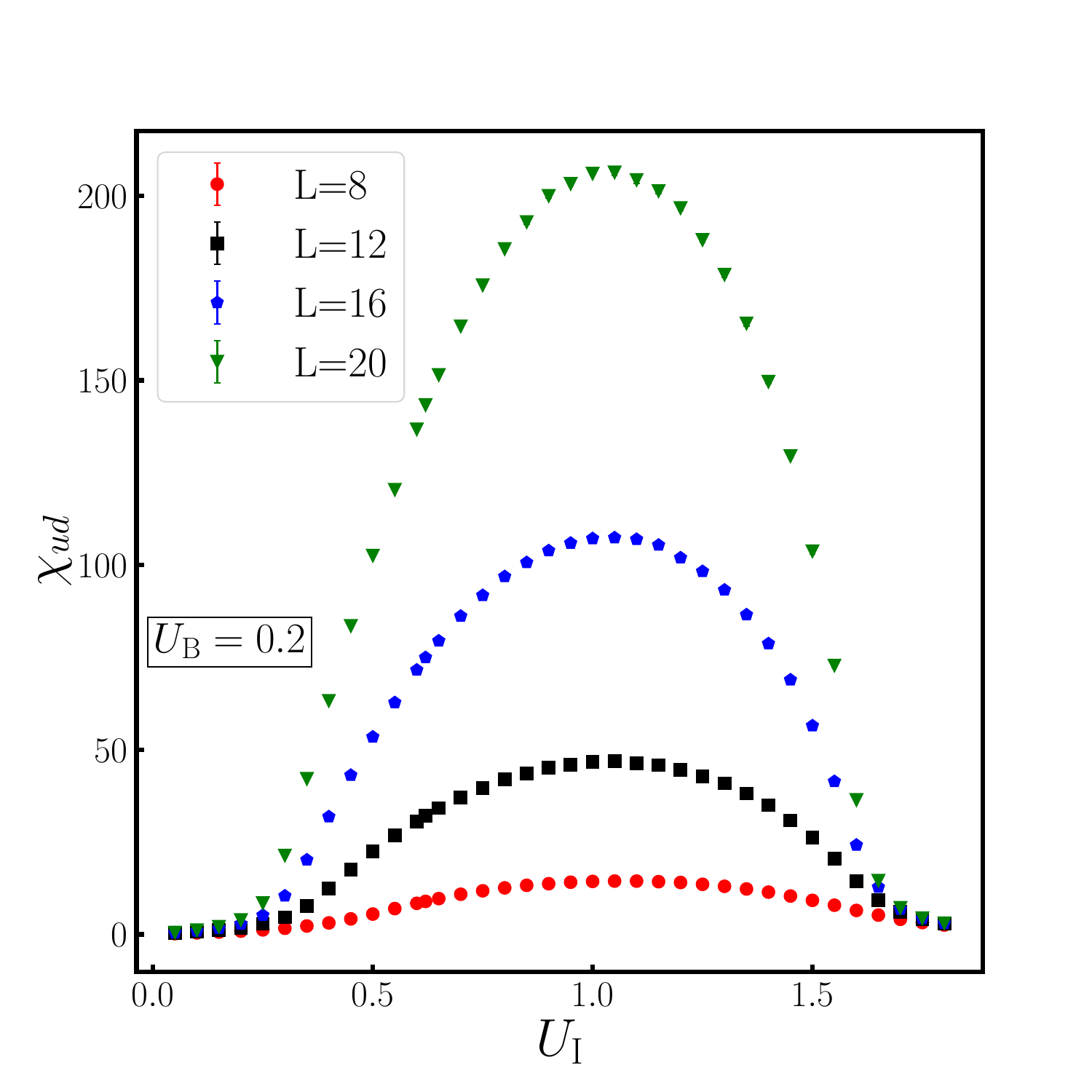}
    \caption{(Top) Heat-map of $\chi_{ud}$ in the $(U_B,U_I)$ plane for lattice sizes $L=12,16,$ and $20$. (Bottom) $\chi_{ud}$ as a function of $U_I$ at fixed $U_B=0.0,\,0.1,$ and $0.2$.}
    \label{fig:hmap}
\end{figure}
To address this, we compute the fermion bilinear susceptibilities using the fermion bag Monte Carlo method. The heat-map of $\chi$ in the $(U_B,U_I)$ plane (top panel of Fig.~\ref{fig:hmap}) reveals three distinct regimes. In the massless and SMG regions, $\chi$ saturates with increasing lattice size, whereas in an intermediate band it is strongly enhanced, signaling spontaneous symmetry breaking. The bottom panel of Fig.~\ref{fig:hmap} shows $\chi$ as a function of $U_I$ for fixed $U_B$, illustrating this behavior across the phase boundaries. Clear signatures of SSB phase persist down to $U_B=0.01$ at $U_I=1.0$; the corresponding volume scaling, $\chi \sim V$, was explicitly demonstrated in Ref.~\cite{Maiti:2026log} by studying the $L$ dependence of $\chi$.
These results support a phase diagram in which any nonzero $U_B$ induces an intermediate SSB phase bounded by two continuous phase transitions. The single transition observed at $U_B=0$ can therefore be interpreted as a multicritical point organizing the surrounding phase structure.

\section{Results}
\label{sec-4}
To determine the nature of the two transitions, we perform finite-size scaling (FSS) of the susceptibilities at fixed $U_B = 0.1$. Near a continuous transition, the data are fitted to
\begin{align}
\chi(U_I,L) = L^{2-\eta}\, f\!\left[(U_I - U_I^c)L^{1/\nu}\right],
\label{eq:eq-fss}
\end{align}
where $f(x)$ is approximated by a polynomial in the scaling variable $x = (U_I - U_I^c)L^{1/\nu}$. Combined fits are performed by constraining $U_I^c$, $\nu$, and $\eta$ to be common to both observables $\chi_{ud}$ and $\chi_{uu}$. Full fit parameters for both the transitions are summarized in Tables~\ref{tab:GN-chifits} and~\ref{tab:XY-chifits}, while further details and consistency checks are presented in Ref.~\cite{Maiti:2026log}.

\begin{figure}
    \centering
    \includegraphics[width=0.42\linewidth, trim=0 0 0 0, clip]{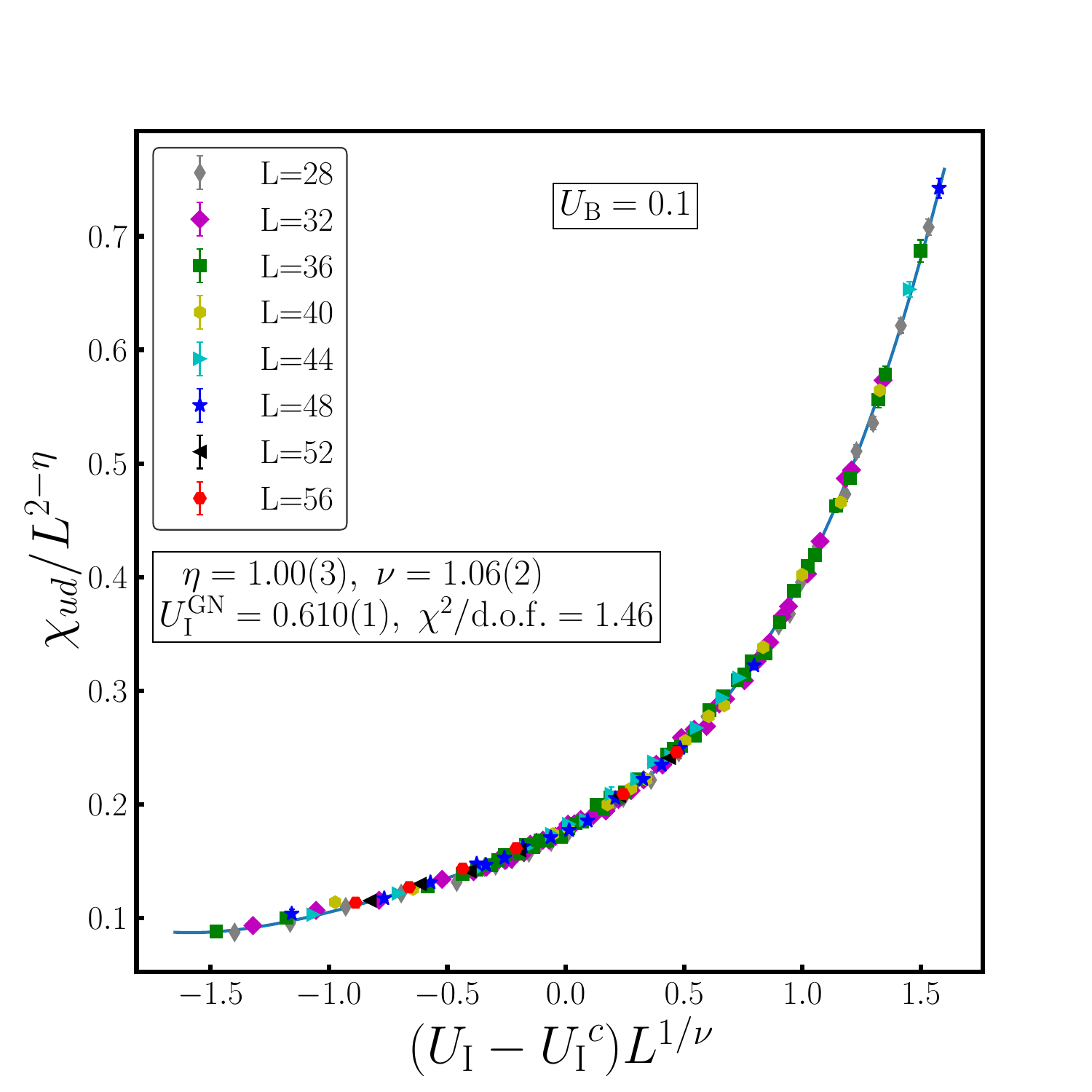}
    \includegraphics[width=0.42\linewidth, trim=0 0 0 0, clip]{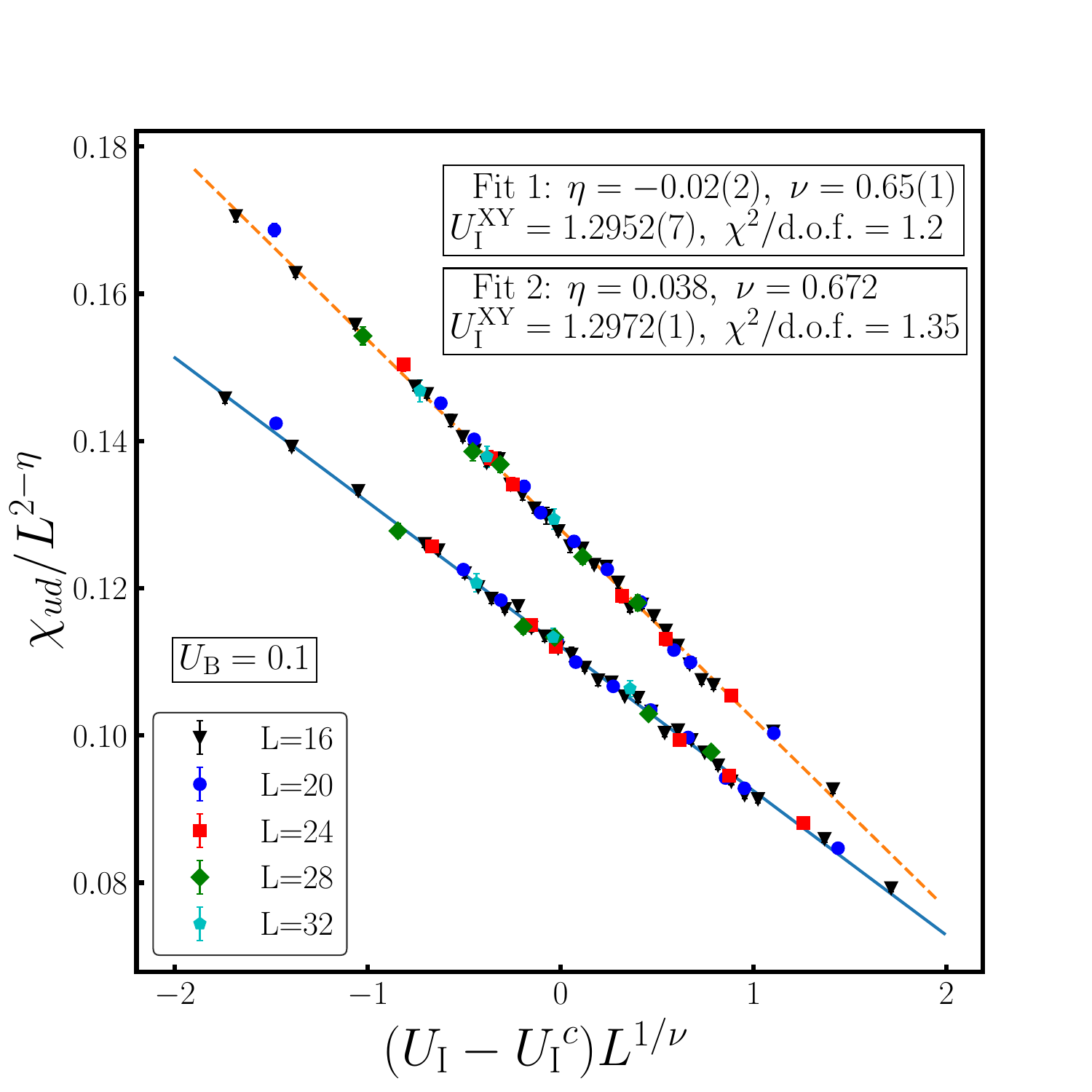}
    \caption{Finite-size scaling collapse from Eq.~\ref{eq:eq-fss} at $U_B=0.1$. 
    (Left): Gross–Neveu transition with a fourth-order polynomial fit and freely varying critical exponents. (Right): 3D–XY transition; the solid line (Fit 1) corresponds to the fit with $\nu$ and $\eta$ treated as fit parameters, while the dashed line (Fit 2) shows the collapse obtained by fixing the exponents to their known 3D–XY values \cite{PhysRevB.63.214503}.}
    \label{fig:FSS}
\end{figure}

\paragraph{MF–SSB phase transition:}
For the transition between the MF and SSB phases, we approximate the scaling function by a fourth-order polynomial. The combined fit yields
\begin{align}
U_I^{\rm GN}=0.610(1), \qquad \nu = 1.06(2), \qquad \eta = 1.00(3).
\end{align}
The corresponding scaling collapse is shown in the left panel of Fig.~\ref{fig:FSS}.
These values are consistent with Gross–Neveu critical behavior in $(2+1)$ dimensions and lie close to the large-$N_f$ mean-field expectations $\nu=1$ and $\eta=1$ \cite{Hands:1992be}. This is natural since two staggered flavors correspond to four continuum Dirac fermions.

\begin{table*}
\renewcommand{\arraystretch}{1.9}
\setlength{\tabcolsep}{2.0pt}    
\centering
\begin{tabular}{c c c c c c c c c c c}
\toprule
& $f_0$ & $f_1$ & $f_2$ & $f_3$ & $f_4$ & $L$-range & ${U_I}^{\rm GN}$ & $\nu$ & $\eta$  & $\chi^2/{\rm d.o.f.}$ \\
\midrule
$\chi_{uu}$ &  0.19(1) & 0.11(1) & 0.060(7) & 0.039(7) & 0.015(4) & 
\multirow{2}{*}{28--56}
& \multirow{2}{*}{0.610(1)}
 & \multirow{2}{*}{1.06(2)} & 
 \multirow{2}{*}{1.00(3)} & \multirow{2}{*}{1.46} \\
$\chi_{ud}$ & 0.18(1)  & 0.11(1) & 0.061(7) & 0.040(7) & 0.014(4) &  &  &  &  & \\
\bottomrule
\end{tabular} 
\caption{Combined fit results for the Gross–Neveu (MF–SSB phase) transition using a fourth-order polynomial scaling function. The critical parameters are common to both $\chi_{uu}$ and $\chi_{ud}$.}
\label{tab:GN-chifits}
\end{table*}

\paragraph{SSB–SMG phase transition:}
For the SSB–SMG phase transition, a linear approximation to the scaling function is sufficient. Allowing the exponents to vary freely gives
\begin{align}
U_I^{\rm XY}=1.2952(7), \qquad \nu = 0.65(1), \qquad \eta = -0.02(2).
\end{align}
The associated data collapse is shown in the right panel of Fig.~\ref{fig:FSS}.
Fixing the exponents to their established 3D–XY values ($\nu=0.672$, $\eta=0.038$) \cite{PhysRevB.63.214503} also leads to a stable scaling collapse, with the critical coupling $U_I^{\rm XY}=1.2972(1)$. This supports the identification of the transition with the 3D-XY universality class, consistent with the decoupling of massive fermions at criticality.
\begin{table*}
\renewcommand{\arraystretch}{1.9}
\setlength{\tabcolsep}{2.0pt}   
\centering
\begin{tabular}{c c c c c c c c}
\toprule
& $f_0$ & $f_1$ & $L$-range & ${U_I}^{\rm XY}$ & $\nu$ & $\eta$  & $\chi^2/d.o.f$ \\
\midrule
$\chi_{uu}$ & 0.113(5) & -0.020(2) & \multirow{2}{*}{16-32}
& \multirow{2}{*}{1.2952(7)}
 & \multirow{2}{*}{0.65(1)} & 
 \multirow{2}{*}{-0.02(2)} & \multirow{2}{*}{1.2} \\
$\chi_{ud}$ & 0.112(5) & -0.020(2) & & & & \\
\midrule
$\chi_{uu}$ & 0.1287(2) & -0.0257(1) & \multirow{2}{*}{16-32}
& \multirow{2}{*}{1.2972(1)}
 & \multirow{2}{*}{0.672} & \multirow{2}{*}{0.038} 
 & \multirow{2}{*}{1.35} \\
$\chi_{ud}$ & 0.1280(2) & -0.0258(1) & & & & \\
\bottomrule
\end{tabular} 
\caption{Combined fit results for the 3D XY transition (SSB-SMG) in the critical region for $\chi_{uu}$ and $\chi_{ud}$ assuming a linear universal scaling function. The first block corresponds to fits where the critical exponents are allowed to vary freely, while the second block uses fixed 3D-XY exponents.}
\label{tab:XY-chifits}
\end{table*}
Taken together, these results demonstrate that, for nonzero $U_B$, the model exhibits two distinct continuous phase transitions: a Gross–Neveu-type transition between MF and SSB phases, and a 3D-XY transition between SSB and SMG phases.

\section{Conclusions}
\label{sec-5}
We have studied the phase structure of a three-dimensional lattice model of two flavors of massless staggered fermions interacting through two independent four-fermion couplings, $U_I$ and $U_B$, using fermion-bag Monte Carlo simulations. Our results show that turning on a nonzero $U_B$ qualitatively changes the phase diagram. While at $U_B=0$ the system exhibits a single continuous transition between the massless and symmetric massive phases, any nonzero $U_B$ generates an intermediate symmetry-broken phase characterized by a fermion bilinear condensate.

Finite-size scaling analysis demonstrates that the MF–SSB phase transition is consistent with Gross–Neveu critical behavior, while the SSB–SMG phase transition belongs to the 3D-XY universality class. These findings support the interpretation that the direct transition observed along the $U_B=0$ axis corresponds to a multicritical point where the Gross–Neveu and 3D-XY critical lines merge. Our study thus provides a coherent lattice picture connecting conventional symmetry-breaking mechanisms and symmetric mass generation within a single framework.


\bibliographystyle{JHEP}
\bibliography{article}

\end{document}